\documentclass[conference]{IEEEtran}
 \usepackage{amsmath,amssymb}
 \usepackage{subfigure}
 \usepackage{graphicx,graphics,color,psfrag}
 \usepackage{cite,balance}
 \usepackage{caption}
 \allowdisplaybreaks
 \usepackage{algorithm}
 \usepackage{accents}
 \usepackage{amsthm}
 \usepackage{bm}
 \usepackage{algorithmic}
 \usepackage[english]{babel}
 \usepackage{multirow}
 \usepackage{enumerate}
 \usepackage{cases}
 \usepackage{stfloats}
 \usepackage{dsfont}
 \usepackage{color,soul}
 \usepackage{amsfonts}
 \usepackage{cite,graphicx,amsmath,amssymb}
 \usepackage{subfigure}
 \usepackage{fancyhdr}
 \usepackage{hhline}
 \usepackage{graphicx,graphics}
 \usepackage{array,color}
 \usepackage{amsmath}
 \usepackage{booktabs}

\newtheorem{lemma}{Lemma}
\newtheorem{corollary}{Corollary}

\newtheorem{proposition}{Proposition}

\def\phi{\varphi}

\def\l{\left}
\def\r{\right}
\def\({\left(}
\def\){\right)}

\def\b0{{\mathbf{0}}}

\begin{document}

\title{\huge Optimizing Wirelessly Powered Crowd Sensing:\\ Trading energy for data}
\author{Xiaoyang Li*\ddag, Changsheng You*, Sergey Andreev\dag, Yi Gong\ddag, and Kaibin Huang*\\
*Dept. of EEE, The University of Hong Kong, Hong Kong \\
\dag Laboratory of Electronics and Communications Engineering, Tampere University of Technology, Finland\\
\ddag Dept. of EEE, Southern University of Science and Technology, Shenzhen, China \\
Email: lixy@eee.hku.hk, csyou@eee.hku.hk, sergey.andreev@tut.fi, gongy@sustc.edu.cn, huangkb@eee.hku.hk
}
\maketitle

\begin{abstract}
To overcome the limited coverage in traditional wireless sensor networks, \emph{mobile crowd sensing} (MCS) has emerged as a new sensing paradigm. To achieve longer battery lives of user devices and incentive human involvement, this paper presents a novel approach that seamlessly integrates MCS with wireless power transfer, called \emph{wirelessly powered crowd sensing} (WPCS), for supporting crowd sensing with energy consumption and offering rewards as incentives. The optimization problem is formulated to simultaneously maximize the data utility and minimize the energy consumption for service operator, by jointly controlling wireless-power allocation at the \emph{access point} (AP) as well as sensing-data size, compression ratio, and sensor-transmission duration at \emph{mobile sensor} (MS). Given the fixed compression ratios, the optimal power allocation policy is shown to have a \emph{threshold}-based structure with respect to a defined \emph{crowd-sensing priority} function for each MS. Given fixed sensing-data utilities, the compression policy achieves the optimal compression ratio. Extensive simulations are also presented to verify the efficiency of the contributed mechanisms.
\end{abstract}

\section{Introduction}
Recently, leveraging a massive number of sensors in pervasive wearable devices and opportunistic human mobility, \emph{mobile crowd sensing} (MCS) has emerged as a new sensing paradigm that includes human as part of the sensing infrastructure \cite{ganti2011mobile}. Among others, prolonging device battery lives and incentivizing human involvement are two key design challenges. One promising solution approach is employing the advanced \emph{wireless power transfer} (WPT) technique for powering \emph{mobile sensor} (MS) in exchange for sensing data, that motivates the current framework, called \emph{wirelessly powered crowd sensing} (WPCS) which is firstly proposed in \cite{galinina2016wirelessly}.

The recent development of MCS has motivated fast-growing research on developing advanced technologies that are more complex than traditional WSN, such as personal location-based systems, trustworthy sensing platforms, and people-centric sensing applications \cite{ganti2011mobile}. One key design issue of MCS is how to involve people in collaborative sensing. There exist two solution approaches, called \emph{opportunistic} and \emph{participatory} sensing \cite{ma2014opportunities}. The former refers to the paradigm where the applications may run in the background and opportunistically collect data without involvement of users, which is fully \emph{unconscious}. While the latter requires users to \emph{consciously} participate the sensing activities initiated by the operator and transmit the required data back timely. This, however, may consume significant resource at the devices (e.g., battery and computing power) and even pose potential privacy threats. Hence, it is essential to design incentive mechanisms to engage the participants with sufficient rewards for compensating the cost, which has been extensively studied in recent papers \cite{zhang2016incentives,he2017exchange,gan2017incentivize}. However, the issue of energy consumption at MSs persists in the prior work. In contrast, the current work resolves this issue by leveraging the advanced WPT techniques.

The WPT technology initially designed for point-to-point power transmission, recently, has been further developed to power wireless transmission by leveraging advanced techniques from inductive coupling and electromagnetic radiation \cite{lu2016wireless}. Combing information and power transmission gives rise to the emergence of an active field called simultaneous wireless information and power transfer (SWIPT) \cite{zhang2013mimo}. A series of research works focus on applying SWIPT to a variety of communication systems and networks, including two-way transmission \cite{popovski2013interactive}, relaying \cite{zheng2017resource}, \emph{multiple-input-multiple-output} (MIMO) communication \cite{zhang2013mimo}, cognitive networks \cite{lee2013opportunistic}, as well as edge-computing networks \cite{you2016energy}. Note that the above existing works focus on optimizing the WPT efficiency for increasing the sensing-and-communication throughput but disregard incentive design and operator's reward. In this work, we integrates WPT with MCS for powering as well as motivating humans to participate in sensing activities.

In this paper, we consider a multiuser WPCS system controlled by an operator that deploys a multiple-antenna \emph{access point} (AP) for transferring energy to multiple single antenna MSs. Each MS stores part of the energy as a reward and applies the rest to operate the sensing tasks including sensing, data compression, and transmission of the compressed data to the AP. We design joint control policy for simultaneously maximizing data utility and minimizing energy consumption at the AP. The main contributions of this work are:

\begin{itemize}
\item \emph{Problem formulation and iterative solution:} To solve the optimization problem, an iterative solution is proposed for the joint optimization of power allocation, sensing, compression, and transmission.
\item \emph{Joint Optimization of Power Allocation, Sensing, and Transmission}: Given the fixed compression ratios, we derive a semi-closed form expression for the optimal sensor-transmission duration. The results are used for deriving the optimal wireless-power allocation and sensing-data sizes. 
\item \emph{Joint Optimization of Compression and Transmission}: Given the fixed sensing-data sizes, the sensor-transmission durations and compression ratios are optimized for minimizing energy consumption at the operator.
\end{itemize}

\section{System model}\label{Sec:Sys}
\begin{figure}[t]
  \centering
  \includegraphics[scale=0.42]{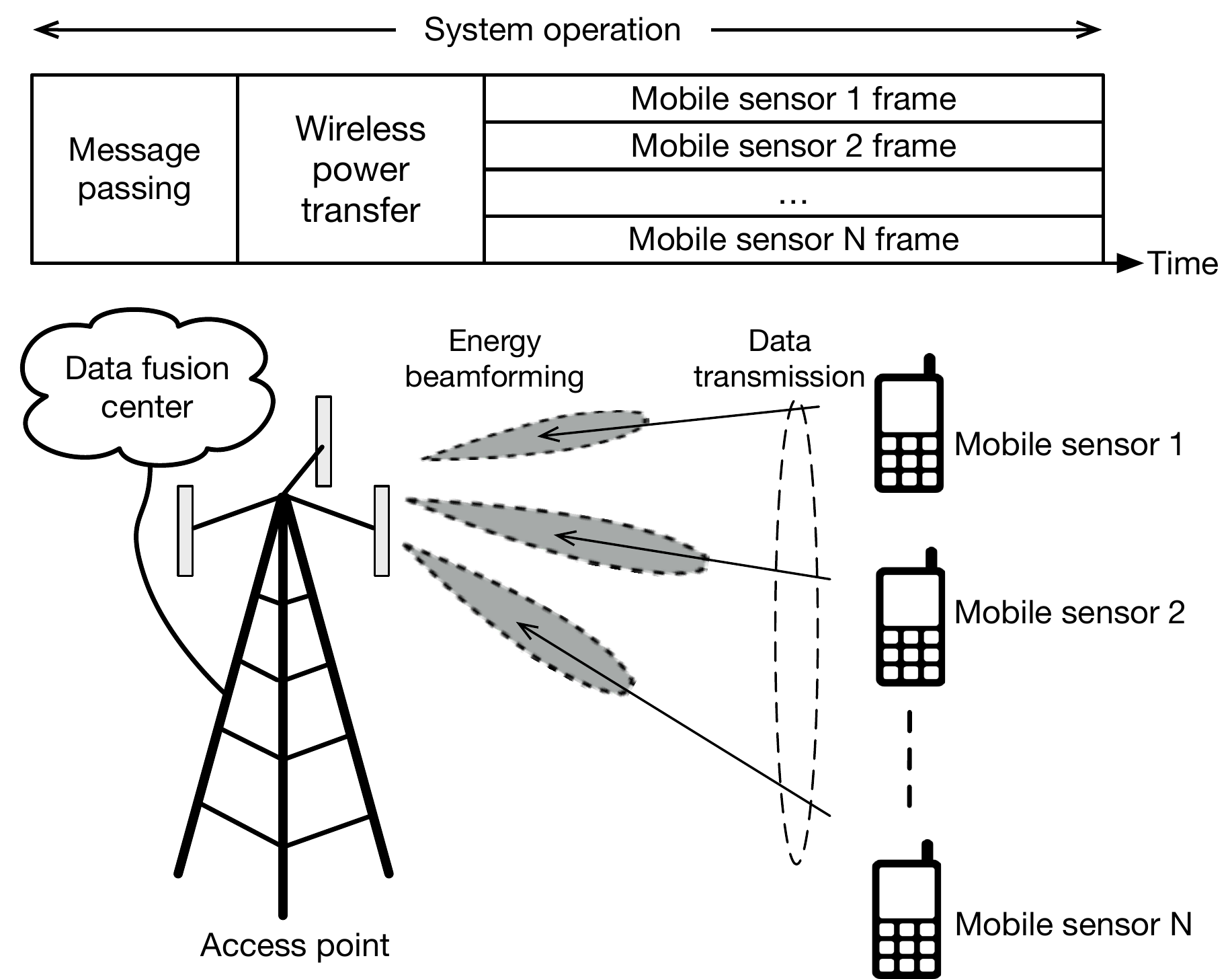}
  \caption{A multiuser WPCS system.}
  \label{FigSys}
\end{figure}

Consider a multiuser WPCS system shown in Fig.~\ref{FigSys} comprising multiple single-antenna MSs and one multi-antenna AP. In \emph{message-passing phase}, the AP jointly optimizes the power allocation and sensor operation based on the feedback parameters from each MS. The parallel data collection is enabled by transmissions over orthogonal channels allocated by the AP to sensors. The time duration for crowd sensing, denoted as $T$, is divided into three slots with durations $t_n^{(s)}$, $t_n^{(c)}$, and $t_n$, which are used for sensing, compression, and transmission, respectively. This introduces the following time constraint:
\begin{equation} \label{Eq:Time:Const}
t_n^{(s)} + t_n^{(c)} + t_n \leq T.
\end{equation}

In a fixed duration of $T_0$ time slots, the AP transfers energy simultaneously to $N$ MSs by multi-antenna beamforming. Let $h_n$ denote the effective channel gain between the AP and MS $n$ and $P_n$ denote the transmit power of the corresponding beam. The AP transmit power is assumed to be fixed and represented by $P_0$, leading to the following power constraint:
\begin{equation} \label{Eq:Power:Const}
\sum_{n=1}^N P_n\leq P_0.
\end{equation}

The energy transferred from the AP to MS $n$, denoted by $E_n^{(h)}(P_n)$, is  $E_n^{(h)}(P_n)=\eta P_n h_n T_0$, where the constant $0<\eta<1$ represents the energy conversion efficiency. Each sensor will store part of the received energy as its reward, denoted by $E_n^{(r)}(\ell_n^{(s)})=q_n^{(r)} \ell_n^{(s)}$ with $q_n^{(r)}$ being a fixed scaling factor. The remaining energy is consumed by sensing, compression, and transmission, represented by $E_n^{(s)}$, $E_n^{(c)}$, and $E_n^{(t)}$, respectively, leading to the following energy constraint: 
\begin{equation}\label{Eq:Energy:Const}
\sum_{n=1}^N \l(E_n^{(r)} + E_n^{(s)} + E_n^{(c)} + E_n^{(t)}\r) \leq \eta h_nP_0 T_0. 
\end{equation}

\begin{figure}[t]
  \centering
  \includegraphics[scale=0.4]{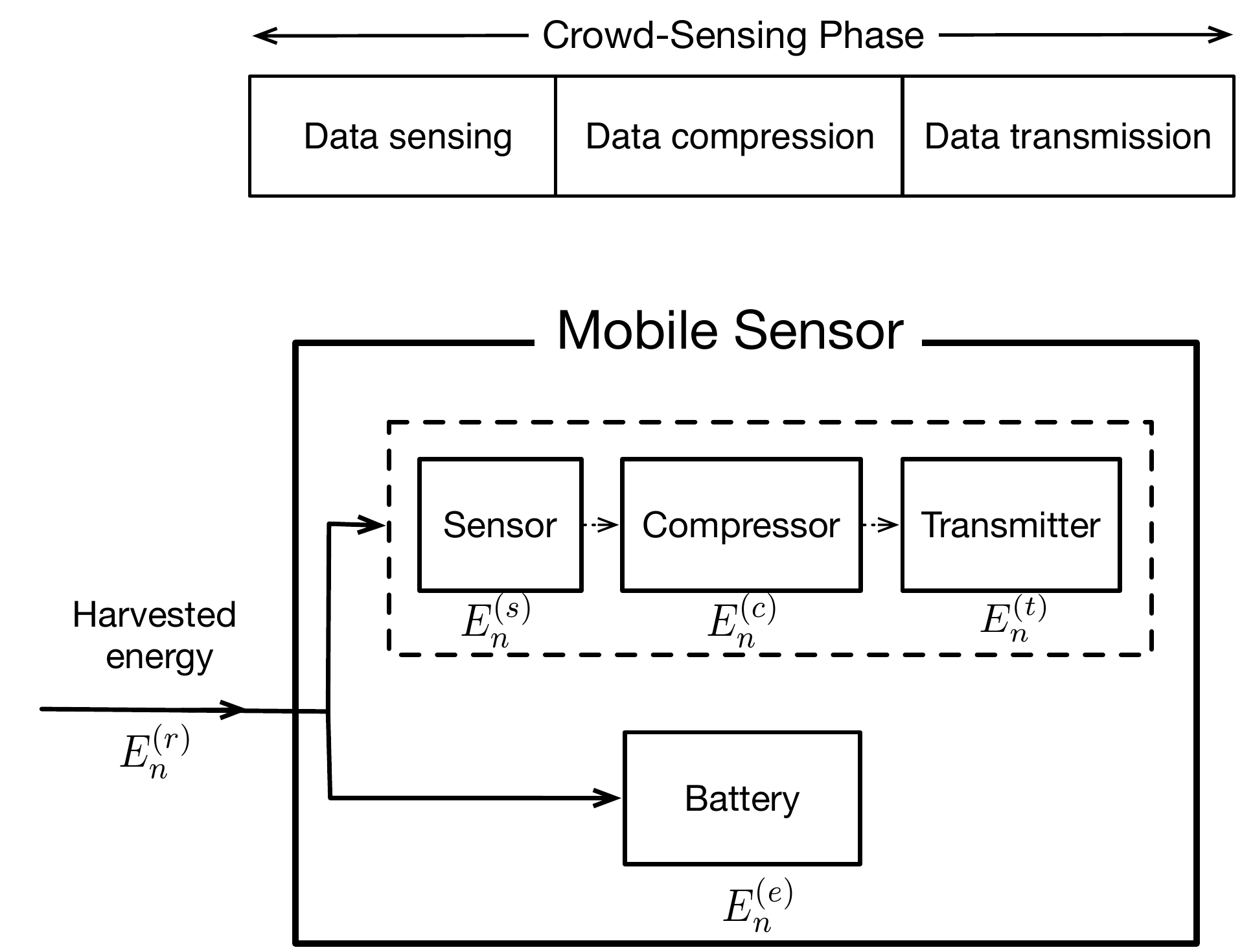}
  \caption{Mobile architecture for WPCS.}
  \label{FigMob}
\end{figure}

At each MS, crowd sensing comprises three sequential operations: data sensing, data compression, and data transmission, as shown in Fig.~\ref{FigMob} and modeled as follows. 

\textbf{\underline{Data Sensing:}} Consider MS $n$. Let $s_n$ denote the output data rate. Given the sensing time duration $t_n^{(s)}$, the size of raw sensing data, denoted by $\ell_n^{(s)}$, is $\ell_n^{(s)}=s_n t_n^{(s)}$. Let $q_n^{(s)}$ denote the sensing energy consumption for generating  $1$-bit of data. The total  energy consumption for sensing at MS $n$, denoted as  $E_n^{(s)}(\ell_n^{(s)})$, is given as $E_n^{(s)}(\ell_n^{(s)})=q_n^{(s)} \ell_n^{(s)}$.

\textbf{\underline{Data Compression:}} Consider lossless compression for sensing data, for which the original data can be perfectly reconstructed from the compressed data. Let $R_{\rm{max}}$ denote the maximum ratio between the sizes of raw and compressed data, thus MS $n$ can choose the compression ratio $R_n\in[1, R_{\rm{max}}]$. Then the size of compressed data is given as  $\ell_n=\ell_n^{(s)}/R_n$. Specifically, the required CPU cycles for compressing $1$-bit of data can be approximated as an exponential function \cite{Arjancompression} of the compression ratio $R_n$ as: 
\begin{equation} \label{Eq:Comp:Complexity}
C(R_n,\epsilon)=e^{\epsilon R_n}-e^{\epsilon},
\end{equation}
where $\epsilon$ is a constant depending on the compression method. Let $f_n$ denote the fixed CPU-cycle frequency at MS $n$, then the compression time duration $t_n^{(c)}=(\ell_n^{(s)} C(R_n,\epsilon))/f_n$. The energy consumption for data compression at MS $n$, denoted by $E_n^{(c)}(\ell_n^{(s)}, R_n)$, is given as $E_n^{(c)}(\ell_n^{(s)}, R_n)=q_n^{(c)} \ell_n^{(s)} C(R_n,\epsilon)$ with $C(R_n,\epsilon)$ in \eqref{Eq:Comp:Complexity}. It follows that 
\begin{equation} \label{Eq:Comp:Energy}
E_n^{(c)}(\ell_n^{(s)}, R_n)=q_n^{(c)} \ell_n^{(s)}\l(e^{\epsilon R_n}\!-\!e^{\epsilon}\r). 
\end{equation}

\textbf{\underline{Data Transmission:}} Each MS transmits its compressed data to the AP. Let $P_n^{(t)}$ denote the transmit power and $t_n$ denote the transmission time duration. Assuming channel reciprocity, the achievable transmission rate (in bit/s) can be given as $v_n=\ell_n/t_n=B\log_2(1+\frac{h_n P_n^{(t)}}{N_0})$, where $B$ is the bandwidth and $N_0$ is the variance of complex-white-Gaussian noise. As such, the transmission energy consumption denoted by $E_n^{(t)}(\ell_n)$ follows: $E_n^{(t)}(\ell_n)=P_n^{(t)} t_n=\dfrac{t_n}{h_n}f(\ell_n/t_n)$, where the function $f(x)$ is defined as $f(x)=N_0(2^{\frac{x}{B}}-1)$.

The \emph{operator's reward} is provided as follows. Following a commonly used model \cite{yang2012crowdsourcing}, the utility of $\ell_n$-bit data delivered by sensor $n$ is measured with the logarithmic function $a_n \log\l(1+\ell_n^{(s)}\r)$, where $a_n$ is a weight factor that depends on the type of data. The sum data utility for the operator can be expressed as
\begin{equation} \label{Eq:Utility}
U(\boldsymbol{\ell}_n^{(s)})=\sum_{n=1}^{N}{a_n}\log(1+b_n \ell_n^{(s)}).
\end{equation}
Then the operator's reward can be modeled as
\begin{align} \label{Eq:Reward}
R(\boldsymbol{\ell}_n^{(s)},P_n)=\sum_{n=1}^{N} a_n \log(1+b_n \ell_n^{(s)})-c\sum_{n=1}^N P_n T_0,
\end{align}
where $c$ denotes the price of unit energy with respect to that of unit data utility.

\section{Problem Formulation and Iterative Solution}
In this section, we formulate and solve the problem of jointly optimizing  power allocation, sensing, lossless compression, and transmission as discussed in Section II. This yields the optimal policy for operating the proposed WPCS system. 

\subsection{Problem Formulation} 
The specific design problem here is to jointly optimize the AP power allocation for WPT to sensors, $\{P_n\}$, the sizes of sensing data, $\{\ell_n^{(s)}\}$, the data compression ratios, $\{R_n\}$, and the partitioning of crowd-sensing time for sensing and compression, determined by $\{t_n\}$ together with $\{\ell_n^{(s)}\}$ and $\{R_n\}$. The objective is to maximize the operator's reward in \eqref{Eq:Reward} under the time constraint in \eqref{Eq:Time:Const}, power constraint in \eqref{Eq:Power:Const}, and energy constraint in \eqref{Eq:Energy:Const}. Mathematically, the optimization problem can be formulated as follows:
\begin{gather*}
\begin{aligned}
\max_{\substack{P_n\ge0, \ell_n^{(s)} \ge0,\\ R_n\in [1,  R_{\rm{max}}], t_n\ge0}} \quad
&\sum_{n=1}^{N} a_n \log(1+\ell_n^{(s)})-c\sum_{n=1}^{N} P_n T_0 \\ 
\textbf{(P1)} \qquad \qquad &\text{s.t.} \qquad \sum_{n=1}^{N} P_n \le P_0,\\
&\frac{\ell_n^{(s)}}{s_n}+\frac{\ell_n^{(s)} C(R_n,\epsilon)}{f_n}+t_n\le T,
\end{aligned}\\
\l[q_n^{(r)}\!+\!q_n^{(s)}\!+\!q_n^{(c)} C(R_n,\epsilon)\r]\ell_n^{(s)}\!+\!\dfrac{t_n}{h_n} f\l(\frac{\ell_n^{(s)}}{t_n R_n}\r)\le \eta P_n h_n T_0.
\end{gather*}

\begin{figure}[t!]
  \centering
  \includegraphics[scale=0.39]{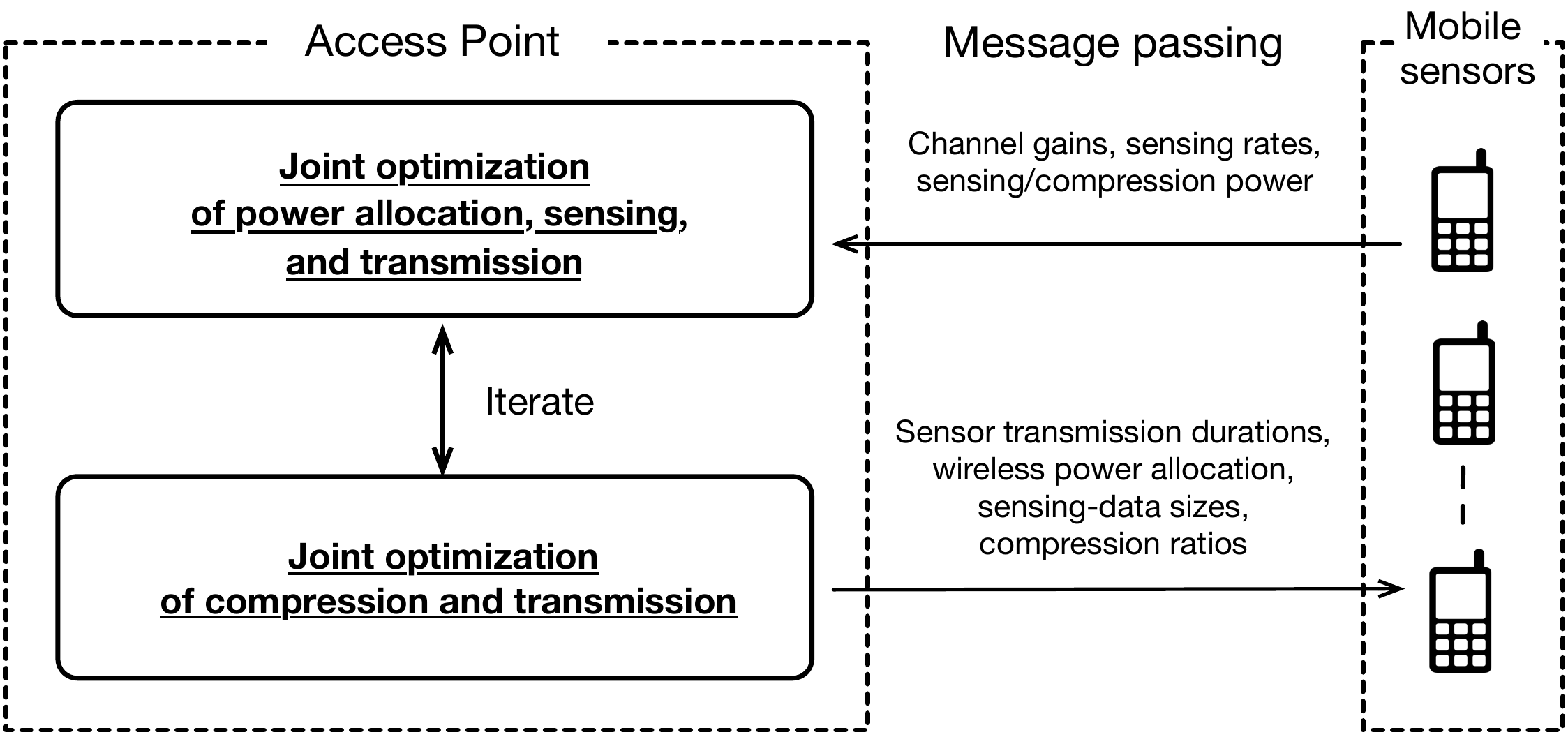}
  \caption{Operator-and-sensor cooperation for WPCS systems.}
  \label{FigNego}
\end{figure}

\subsection{Iterative Solution Approach} \label{Sec:Iterative}
Problem P1 is a non-convex problem, since the optimization variables, $\l\{\ell_n^{(s)}, R_n, t_n\r\}$, are coupled at the constraints in form of  $\ell_n^{(s)} C(R_n,\epsilon)$ and $\dfrac{t_n}{h_n}f\l(\ell_n^{(s)}/(t_n R_n)\r)$. To solve this challenging problem and characterize the policy structure, we propose an \emph{iterative} solution approach as shown in Fig.~\ref{FigNego}. Specifically, it decomposes Problem P1 into two sub-problems, P1-A and P1-B as formulated in the sequel. By iteratively solving these two problems, the solution is guaranteed to converge and reach the local optimum of Problem P1.

\subsubsection{Joint optimization of wireless power allocation, sensing, and transmission}
Given compression ratios $\boldsymbol{R}=[R_1, R_2, \cdots, R_N]$ as fixed, Problem P1 can be reduced to the optimization problem below, which jointly optimizes the wireless power allocation, sensing-data size, and transmission duration for the maximum operator's reward.
\begin{gather*}
\begin{aligned}
\max_{P_n\ge0, \ell_n^{(s)} \ge0, t_n\ge0} \quad 
&\sum_{n=1}^{N}a_n\log(1+\ell_n^{(s)})-c\sum_{n=1}^{N} P_n T_0\\ 
\textbf{(P1-A)} \qquad  \qquad &\text{s.t.} \qquad \sum_{n=1}^{N} P_n \le P_0,\\
&\beta_n \ell_n^{(s)}+t_n\le T,\\
&\alpha_n \ell_n^{(s)}+\dfrac{t_n}{h_n} f\l(\frac{\ell_n^{(s)}}{R_n t_n}\r)\le \eta P_n h_n T_0,
\end{aligned}
\end{gather*}
where $\alpha_n\!=\!q_n^{(r)}\!+\!q_n^{(s)}\!+\!q_n^{(c)} C(R_n,\epsilon)$ and $\beta_n\!=\!\dfrac{1}{s_n}\!+\!\dfrac{C(R_n,\epsilon)}{f_n}$.

\subsubsection{Joint optimization of compression and transmission}
Given the fixed sensing-data size, $\boldsymbol{\ell}^{(s)}$, which results in constant sum data utility, Problem P1 is reduced to minimizing the sum energy consumption: $\sum_{n=1}^{N} P_n T_0$. Moreover, it can be derived from the energy constraints in \eqref{Eq:Energy:Const} that the sum energy consumption can be achieved by separately minimizing $P_n$ for each MS, which should satisfy the following:
\begin{equation}
\l[q_n^{(r)}\!+\!q_n^{(s)}\!+\!q_n^{(c)} C(R_n,\epsilon)\r]\ell_n^{(s)}\!+\!\dfrac{t_n}{h_n} f\l(\frac{\ell_n^{(s)}}{t_n R_n}\r)= \eta P_n h_n T_0.
\end{equation}
Hence, the solution for the minimum sum energy consumption can be derived by solving the following problem for each MS:
\begin{gather*}
\begin{aligned}
\min_{1 \!\le\! R_n \!\le\! R_{\rm{max}}, t_n \!\ge\! 0} 
&\l[q_n^{(r)}\!+\!q_n^{(s)}\!+\!q_n^{(c)} C(R_n,\epsilon)\r]\ell_n^{(s)}\!+\!\dfrac{t_n}{h_n} f\l(\frac{\ell_n^{(s)}}{t_n R_n}\r)\\
\textbf{(P1-B)} ~ \text{s.t.} ~
&\frac{\ell_n^{(s)}}{s_n}+\frac{\ell_n^{(s)} C(R_n,\epsilon)}{f_n}+t_n\le T.
\end{aligned}
\end{gather*}

\section{Joint optimization of wireless power allocation, sensing, and transmission}
In this section, given the fixed compression ratios, we equivalently transform the original optimization problem into the one focusing on the sensor-transmission optimization. Observe that Problem P1-A is always feasible since $\{P_n=0, \ell_n^{(s)}=0, R_n=1\}$ for each $n$ is a feasible solution. To characterize the policy structure, we first present two necessary conditions for the optimal solution in the lemma below, proved in Appendix~\ref{App:NecCond}.

\begin{lemma}\label{Lem:NecCond}
\emph{Given the fixed compression ratios at sensors, the optimal sensor transmission, wireless power allocation, and sensing-data size for each MS solving Problem P1-A, denoted by $t_n^*, P_n^*$, and  $(\ell_n^{(s)})^*$, respectively, satisfy the following:
\begin{gather*}
\alpha_n \l(\ell_n^{(s)}\r)^*+\dfrac{t_n^*}{h_n} f\l(\frac{(\ell_n^{(s)})^*}{R_n t_n^*}\r)=\eta P_n^* h_n T_0,\notag \\
\beta_n \l(\ell_n^{(s)}\r)^*+t_n^*= T.\label{NecCond}
\end{gather*}}
\end{lemma}

It means that to maximize the operator's reward, each MS should \emph{fully} utilize the allocated transferred energy and the crowd-sensing time duration. Using Lemma~\ref{Lem:NecCond}, Problem P1-A can be equivalently transformed into Problem P2 as follows.
\begin{gather*}
\begin{aligned}
\max_{t_n\ge0}~
&\sum_{n=1}^{N}a_n\log\l(1+\frac{T-t_n}{\beta_n}\r)-c\sum_{n=1}^N E(t_n)\\ 
\textbf{(P2)} ~ \text{s.t.} ~
&\sum_{n=1}^N \l[\frac{\alpha_n(T-t_n)}{\eta \beta_n h_n}+\dfrac{t_n}{\eta h_n^2} f\l(\frac{T-t_n}{R_n \beta_n t_n}\r)\r]\le P_0 T_0,\\
&0\le t_n\le T.
\end{aligned}
\end{gather*}
where $E(t_n)=\frac{\alpha_n(T-t_n)}{\eta \beta_n h_n}+\dfrac{t_n}{\eta h_n^2} f\l(\frac{T-t_n}{{R_n}{\beta_n}{t_n}}\r)$. The convexity of Problem P2 is given as following, proved in Appendix~\ref{App:conv}.

\begin{lemma}\label{Lem:conv}
\emph{Problem P2 is a convex optimization problem.}
\end{lemma}

As a result, Problem P2 can be solved by the Lagrange method. The corresponding partial Lagrange function is 
\begin{gather}
L(t_n,\lambda)\!=\!\sum_{n=1}^{N} (\lambda\!+\!c)\l(\frac{\alpha_n(T\!-\!t_n)}{\eta \beta_n h_n}\!+\!\dfrac{t_n}{\eta h_n^2} f\l(\frac{T\!-\!t_n}{R_n \beta_n t_n}\r)\r) \notag \\
\!-\!a_n\log\l(1\!+\!\frac{T\!-\!t_n}{\beta_n}\r)\!-\!\lambda P_0 T_0.
\end{gather}

Define a function $y(x)$ as $y(x)=f(x)-(x+\frac{1}{R_n \beta_n})f'(x)$ and $t_n^*$  as the optimal solution for solving Problem P2. Then applying \emph{Karush-Kuhn-Tucker} (KKT) conditions leads to the following necessary and sufficient conditions:
\begin{gather}
\frac{\partial L}{\partial t_n^*}=\frac{a_n}{\beta_n\!+\!(T\!-\!t_n^*)}\!+\!\frac{(\lambda^*\!+\!c_n)}{\eta h_n}\l[\frac{1}{h_n}y\l(\frac{T-t_n^*}{R_n \beta_n t_n^*}\r)\!-\!\frac{\alpha_n}{\beta_n}\r] \notag\\
\begin{cases}
>0,&t_n^*=0,\\
=0,&0<t_n^*<T,\\
<0,&t_n^*=T,
\end{cases}\label{Eq:KKT1}\\
\lambda^*\l(\sum_{n=1}^{N}\l[\frac{t_n^*}{\eta h_n^2}f\l(\frac{T\!-\!t_n^*}{R_n \beta_n t_n^*}\r)\!+\!\frac{\alpha_n(T\!-\!t_n^*)}{\eta \beta_n h_n}\r]\!-\!P_0 T_0\r)\!=\!0.\label{Eq:KKT2}
\end{gather}
Combining these conditions yields the optimal sensor transmission policy given as follows.

\begin{proposition}[Optimal Sensor Transmission]\label{Pro:LosslessSensorT}\emph{Given the fixed compression ratios, for each MS selected for crowd sensing with $(0<t_n^*\le T)$, the optimal transmission duration $t_n^*$ satisfies:
\begin{gather*}
\frac{(\lambda^*+c) N_0}{\eta h_n^2}\l[\l(1-\frac{T\ln2}{B R_n \beta_n t_n^*}\r)2^\frac{T-t_n^*}{B R_n \beta_n t_n^*}-1\r] \notag\\
+\frac{a_n}{\beta_n+(T-t_n^*)}-\frac{(\lambda^*+c)\alpha_n}{\eta h_n \beta_n}=0,\label{Eq:LosslessOpt}
\end{gather*}
where $\lambda^*$ satisfies the condition in \eqref{Eq:KKT2}.
}
\end{proposition}

One can observe from Proposition~\ref{Pro:LosslessSensorT} that the optimal sensor-transmission duration has no closed form. To derive the values, we propose an efficient bisection method to compute the policy as described in Algorithm~\ref{Al:Transmission}. Based on Proposition~\ref{Pro:LosslessSensorT}, the effects of parameters on the sensor-transmission duration are characterized in the following corollary, proved in Appendix~\ref{App:StruTime}.

\begin{corollary}[Properties of Optimal Sensor Transmission]\label{Cor:StruTime}\emph{The optimal sensor-transmission durations $\{t_n^*\}$ have the following properties:  
\begin{itemize}
\item[\emph{1)}] If the channel gains $\{h_n\}$ are identical but the utility weights satisfy  $a_1\ge a_2\cdots \ge a_N$, then $t_1^*\le t_2^* \cdots \le t_N^*$.
\item[\emph{2)}] If the utility weights $\{a_n\}$ are identical but the channel gains satisfy   $h_1\ge h_2\cdots  \ge h_N$, then $t_1^*\le t_2^* \cdots \le t_N^*$. 
\end{itemize} }
\end{corollary}

These cases indicate that for a sensor with a large utility weight and channel gain, it is beneficial to have a short transmission duration and thus it can take a long duration for sensing larger amounts of data to increase the sum data utility.

\begin{algorithm}[t!]
  \caption{A Bisection Algorithm for Computing the Optimal Sensor Transmissions}
  \label{Al:Transmission}
  \begin{itemize}
\item{\textbf{Step 1} [Initialize]: Let $\lambda_{\ell}=0$, $\lambda_h=\lambda_{\rm{max}}$, $\varepsilon >0$.\\
Obtain $t_{n,\ell}$ and $t_{n,h}$ based on \eqref{Eq:LosslessOpt}, calculate $E_{\ell}=\sum_{n=1}^N \frac{\alpha_n(T-t_{n,\ell})}{\eta \beta_n h_n}+\dfrac{t_{n,\ell}}{\eta h_n^2} f\l(\frac{T-t_{n,\ell}}{R_n \beta_n t_{n,\ell}}\r)$ and $E_h = \sum_{n=1}^N \frac{\alpha_n(T-t_{n,h})}{\eta \beta_n h_n}+\dfrac{t_{n,h}}{\eta h_n^2} f\l(\frac{T-t_{n,h}}{R_n \beta_n t_{n,h}}\r)$, respectively.}
\item{\textbf{Step 2}  [Bisection search]: \emph{While} $E_{\ell} \neq P_0 T_0$ and $E_h \neq P_0 T_0$, update $\lambda_{\ell}$ and $\lambda_h$ as follows.\\
(1) Define $\lambda_m\!=\!(\lambda_{\ell}\!+\!\lambda_h)/2$, compute $t_{n,m}$ and $E_m\!=\!\sum_{n=1}^N \frac{\alpha_n(T\!-\!t_{n,m})}{\eta \beta_n h_n}\!+\!\dfrac{t_{n,m}}{\eta h_n^2} f\l(\frac{T\!-\!t_{n,m}}{R_n \beta_n t_{n,m}}\r)$.\\
(2) If $E_m<P_0 T_0$, let $\lambda_h=\lambda_m$, else $\lambda_{\ell}=\lambda_m$.\\
\emph{Until} $\lambda_h-\lambda_{\ell}<\varepsilon$. Return $t_n^*=t_{n,m}$.}
\end{itemize}
  \end{algorithm}

Based on Proposition~\ref{Pro:LosslessSensorT}, the optimal wireless-power allocation policy is shown as following, proved in Appendix~\ref{App:LosslessPowerAllo}. 

\begin{proposition}[Optimal Wireless-Power Allocation]\label{Pro:LosslessPowerAllo}\emph{Given the fixed compression ratios, the optimal wireless-power allocation policy for maximizing the operator's reward, $P_n^*$, has the following structure:
\begin{equation*}\label{Eq:OptPower}
P_n^*=
\begin{cases}
\dfrac{1}{h_n \eta T_0}\l[\dfrac{t_n^*}{h_n}f\l(\dfrac{T\!-\!t_n^*}{R_n \beta_n t_n^*}\r)\!+\!\dfrac{\alpha_n (T\!-\!t_n^*)}{\beta_n}\r],&\phi_n \! \geq \!\lambda^*,\\
0,&\phi_n \!<\! \lambda^*,
\end{cases}
\end{equation*}
where $t_n^*$ and $\lambda^*$ are given in Proposition~\ref{Pro:LosslessSensorT}, and $\phi_n$ is defined as the \emph{crowd-sensing priority function}, offered by
\begin{equation*}\label{Eq:LosslessPrio}
\phi_n\!=\!\kappa_n \!-\!c~\text{with}~\kappa_n \overset{\triangle}{=}\frac{a_n \eta h_n}{q_n^{(r)}\!+\!q_n^{(s)}\!+\!q_n^{(c)} C(R_n,\epsilon)\!+\!\frac{N_0\ln2}{h_n B R_n}}.
\end{equation*}}
\end{proposition}

Proposition~\ref{Pro:LosslessPowerAllo} reveals that given the information of MSs, the optimal wireless-power allocation policy has a \emph{threshold}-based structure. In other words, only the MSs with (crowd-sensing) priority functions exceeding the threshold $\lambda^*$ will be allocated with wireless power for participating in the crowd sensing.

\section{Joint optimization of compression and transmission}
Consider the optimization problem formulated in Problem P1-B. First, the problem convexity is established in the following lemma, which can be easily proved by using the property of \emph{perspective function} \cite{Boyd2006convex}..
\begin{lemma}\label{Lem:losslessconv}
\emph{Problem P1-B is a convex optimization problem.}
\end{lemma}

Therefore, Problem P1-B can be solved by the Lagrange method. The partial Lagrange function is thus
\begin{gather}
\hat{L}(R_n,t_n,\hat{\lambda})=Q(R_n)\ell_n^{(s)}+\dfrac{t_n}{h_n}f\l(\frac{\ell_n^{(s)}}{t_n R_n}\r) \notag\\
+\hat{\lambda}\l(t_n+\frac{\ell_n^{(s)}C(R_n,\epsilon)}{f_n}+\frac{\ell_n^{(s)}}{s_n}-T\r),
\end{gather}
where $Q(R_n)=q_n^{(r)}+q_n^{(s)}+q_n^{(c)}C(R_n,\epsilon)$. Let $g(x)=f(x)-x f^{'}(x)$ and $t_n^*$, $R_n^*$ denote the optimal solution for solving Problem P1-B. Directly applying the KKT conditions results in the following necessary and sufficient conditions:
\begin{gather}\label{Eq:LosslessKKT}
\frac{\partial \hat{L}}{\partial R_n^*}=\l(q_n^{(c)}+\frac{\hat{\lambda}^*}{f_n}\r)\ell_n^{(s)} \epsilon e^{\epsilon R_n^*}-\frac{\ell_n^{(s)}}{h_n R_n^{*2}}f'\l(\frac{\ell_n^{(s)}}{t_n^* R_n^*}\r) \notag\\
\begin{cases}
>0,&R_n^*=1,\\
=0,&1<R_n^*<R_{\rm{max}},\\
<0,&R_n^*=R_{\rm{max}},
\end{cases}\label{Eq:LosslessKKT1}\\
\frac{\partial{\hat{L}}}{\partial{t_n^*}}=\frac{1}{h_n}g\l(\frac{\ell_{n}^s}{t_n^* R_n^*}\r)+{\hat{\lambda}}^{*}
\begin{cases}
>0, &t_n^*=0,\\
=0, &0<t_n^*<T,\\
<0, &t_n^*=T,
\end{cases}\label{Eq:LosslessKKT2}\\
\hat{\lambda}^*\l(t_n^*+\frac{\ell_n^{(s)}C(R_n,\epsilon)}{f_n}+\frac{\ell_n^{(s)}}{s_n}-T\r)=0.\label{Eq:LosslessKKT3}
\end{gather}

Based on the conditions in \eqref{Eq:LosslessKKT1}-\eqref{Eq:LosslessKKT3}, the key result of this subsection is derived, stated as the following, proved in Appendix~\ref{App:LosslessDataComp}.

\begin{proposition}[Optimal Compression and Transmission]\label{Pro:LosslessDataComp}\emph{The solution of Problem P1-B is as follows. 
\begin{itemize}
\item[1)] The optimal sensor-transmission durations $\{t_n^*\}$ are
\begin{equation}
t_n^*= T -\dfrac{\ell_n^{(s)}\l(e^{\epsilon R_n^*}-e^{\epsilon}\r)}{f_n} -\dfrac{\ell_n^{(s)}}{s_n}, \qquad \forall \ n,
\end{equation}
which corresponds to full utilization of the crowd-sensing duration.
\item[2)] The optimal compression ratios $\{R_n^*\}$ are
\begin{equation}
R_n^*=\max\l\{\min\l\{\hat{R}_n, R_{\rm{max}}\r\}, 1\r\}, \qquad \forall \ n,
\end{equation} where $\hat{R}_n$ satisfies: $z(\hat{R}_n)=0$ with the function $z(R_n)$ defined by
\begin{gather}\label{Eq:LosslessR}
z(R_n)=\l[q_n^{(c)}-\frac{1}{h_n f_n}g\l(\frac{1}{d(R_n)R_n}\r)\r]\epsilon e^{\epsilon R_n} \notag\\
-\frac{1}{h_n R_n^2}f'\l(\frac{1}{d(R_n)R_n}\r),
\end{gather}
and $d(R_n)\overset{\triangle}{=}\dfrac{T}{\ell_n^{(s)}}-\dfrac{1}{s_n}-\dfrac{1}{f_n}C(R_n,\epsilon)>0$.
\end{itemize}}
\end{proposition}

Even though $\hat{R}_n$ has no closed form, it can be computed by the bisection method as detailed in Algorithm~\ref{Al:OptLossless}. This is due to the monotone property of function $z(R_n)$ proved in Appendix~\ref{App:Losslessmonotone}.

\begin{algorithm}[t]
  \caption{Optimal Lossless Compression Ratio Algorithm}
  \label{Al:OptLossless}
  \begin{itemize}
\item{\textbf{Step 1} [Initialize]: Let $R_{\ell}=1$, $R_h=R_{\max}$, $\varepsilon >0$. Obtain $z(R_{\ell})$, $z(R_h)$.}
\item{\textbf{Step 2} [Bisection search]: \emph{While} $z(R_{\ell}) \neq 0$ and $z(R_h) \neq 0$, update $R_{\ell}$ and $R_h$ as follows.\\
(1) Define $R_n^*=(R_{\ell}+R_h)/2$, compute $z(R_n^*)$.\\
(2) If $z(R_n^*)>0$, let $R_h=R_n^*$, else $R_{\ell}=R_n^*$.\\
\emph{Until} $R_h-R_{\ell}<\varepsilon$.}
\end{itemize}
  \end{algorithm}

\begin{lemma}\label{Lem:Losslessmonotone}
\emph{$z(R_n)$ is a \emph{monotone-increasing} function of $R_n$.}
\end{lemma}

Further, the effects of parameters on the optimal compression ratio are detailed in the following, proved in Appendix~\ref{App:CompressionRatio}.

\begin{corollary}[Properties of Optimal Compression Ratios]\label{Cor:CompressionRatio}\emph{The optimal compression ratios $\{R_n^*\}$ have the following properties:
\begin{itemize}
\item[\emph{1)}] If compression powers $\{q_n^{(c)}\}$ are identical but channel gains satisfy $h_1\!\ge \!h_2\!\cdots \!\ge\! h_N$, then $R_1^*\!\le\! R_2^*\! \cdots\! \le\! R_N^*$.
\item[\emph{2)}] If channel gains $\{h_n\}$ are identical but compression powers satisfy $q_1^{(c)}\!\ge\! q_2^{(c)}\! \cdots \! \ge\! q_n^{(c)}$, then $R_1^*\!\le\! R_2^* \!\cdots \!\le\! R_N^*$.
\end{itemize} }
\end{corollary}


Combining the results in the preceding sections, the efficient control policy for achieving the maximum operator's reward can be computed by an iterative algorithm summarized in Algorithm~\ref{Al:Lossless}.


\begin{algorithm}[t]
  \caption{Efficient Algorithm for Lossless Compression Solving Problem P1}
  \label{Al:Lossless}
  \begin{itemize}
\item{\textbf{Step 1}: Initialize the compression ratio $R_n^*$.}
\item{\textbf{Step 2}: \emph{Repeat} \\
(1) Given the fixed compression ratio, find the optimal power allocation $P_n^*$ and the sensing-data size $(\ell_n^{(s)})^*$ by using Algorithm~\ref{Al:Transmission}.\\
(2) Given the fixed sensing-data size, update the compression ratio $R_n^*$ by using Algorithm~\ref{Al:OptLossless}.\\
\emph{Until} convergence or a maximum number of iterations has been reached.}
  \end{itemize}
  \end{algorithm}

\section{Numerical Results and Discussion}
In this section, the performance of proposed operator-and-mobile cooperation algorithm is evaluated by simulations. The WPCS system comprises $1$ AP and $10$ MSs where the AP transfers wireless power to MSs in the time duration $T_0=1$ s. The channels $\{h_n\}$ are modeled as independent Rayleigh fading with average signal attenuation set as $10^{-3}$. The bandwidth is $B_n=10$ KHz and the variance of complex-white-Gaussian-channel noise is $N_0=10^{-9}$ W. For the system reward, the utility weights of all the MSs are $a_n=0.04$ and the cost weight is $c=0.6$. For each MS, the sensing rate follows a uniform distribution with $s_n \in [1,10] \times 10^4$ bit/s with energy consumption per bit distributed by $q_n^{(s)} \in [1,10] \times 10^{-12}$ J/bit. For the data compression, the required number of CPU cycles is $C_n \in [0,3000]$ cycles/bit, CPU-cycle frequency is $f_n \in [0.1,1]$ GHz, and energy consumption per cycle is $q_n^{(c)} \in [1,10] \times 10^{-14}$ J/bit, respectively. In addition, the energy reward per bit is uniformly distributed as: $q_n^{(r)} \in [1,10] \times 10^{-12}$ J/bit. The maximum compression ratio is $R_{\max} = 3$ and $\epsilon = 4$.

Two baseline policies are considered for performance comparison. The first one optimizes the power allocation with the \emph{fixed compression ratio} (FCR). Specifically, the fixed compression ratio is set as $R_n=1.5$. The second policy considers no compression operation. 

\begin{figure}[t]
  \centering
  \includegraphics[scale=0.35]{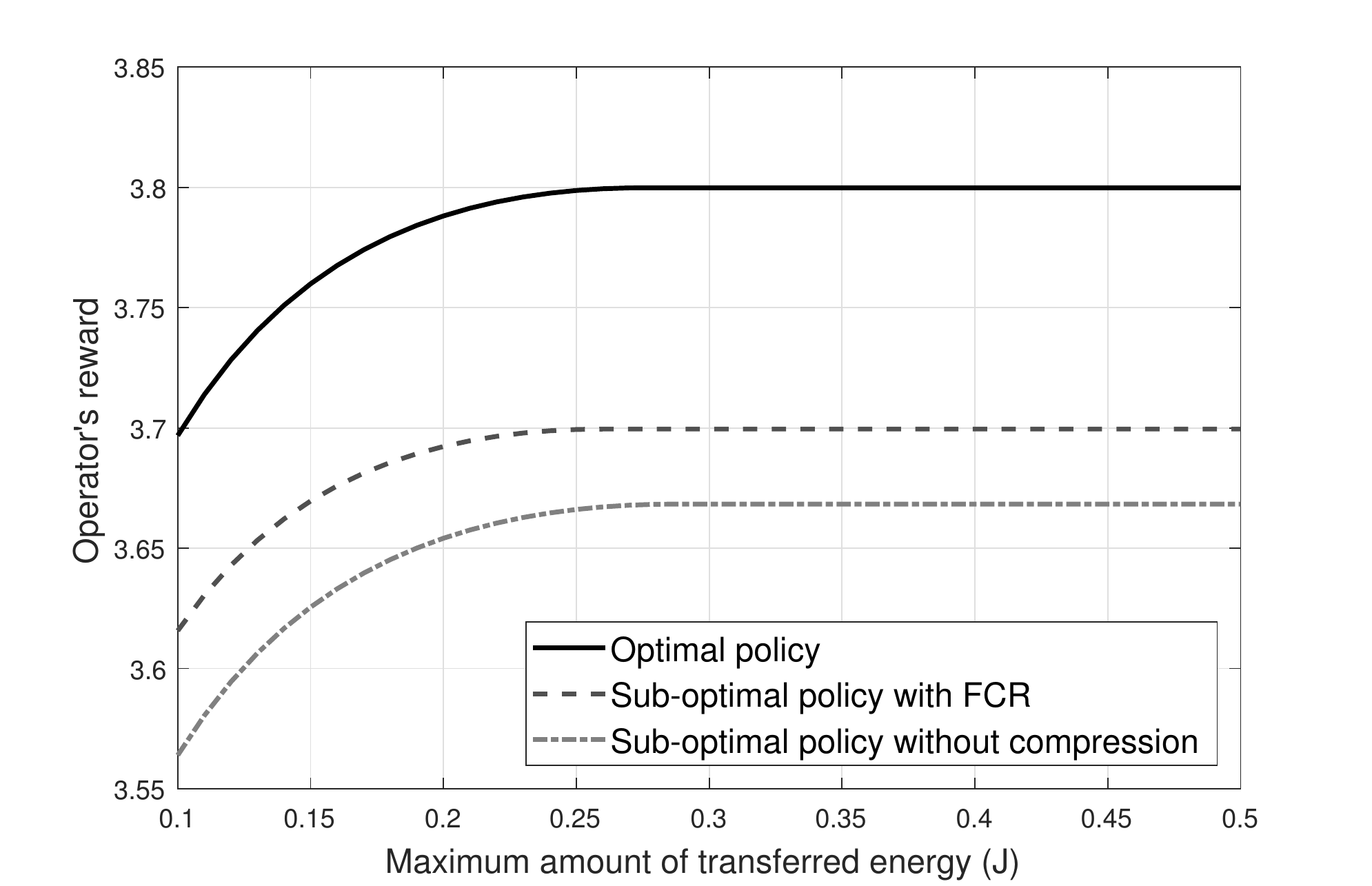}
  \caption{Operator's reward versus the maximum transferred energy.}
  \label{FigRewardE}
\end{figure}

The values of the operator's reward versus the maximum amount of transferred energy are displayed in Fig.~\ref{FigRewardE}. The time duration for crowd-sensing is set as $T=1$ s. It can be observed that the operator's reward for the optimal policy is log-like and increases with the transferred energy. The optimal policy has significant performance gain over the two baseline policies.

\begin{figure}[t]
  \centering
  \includegraphics[scale=0.35]{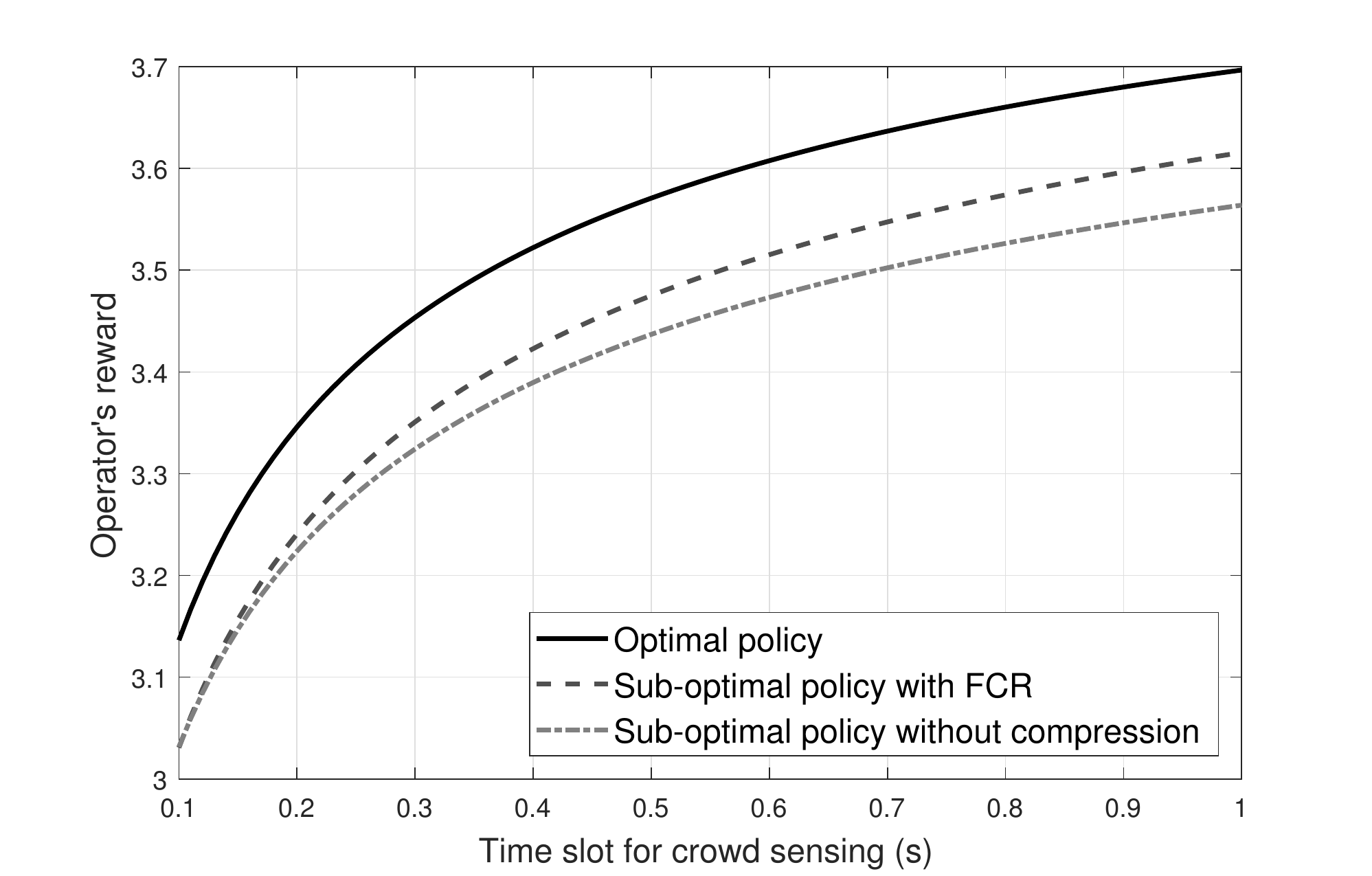}
  \caption{Operator's reward versus the time slot for crowd sensing.}
  \label{FigRewardT}
\end{figure}

Further, Fig.~\ref{FigRewardT} shows the values of the operator's reward versus the crowd-sensing time duration for both compression methods. It can be observed that the operator's reward is monotone-increasing with an extension of crowd-sensing time duration, as it can reduce transmission-energy consumption so as to spare more energy for data sensing. 

\section{Concluding Remarks}
This work studied power allocation for a multi-user WPCS system and the optimal compression adjustment. A crowd-sensing priority function was derived to select the MS. An iterative algorithm was designed to maximize the operator's reward, which has the optimal performance in simulations. This work can be further extended to other interesting scenarios such as lossy compression or distributed inference.

\section*{Acknowledgment}
This work was supported in part by Hong Kong Research Grants Council under the Grants 17209917 and 17259416, Natural Science Foundation of China under Grant 61531009, and Natural Science Foundation of Guangdong Province under Grant 2015A030313844.

\begin{appendices}
\section{Proof of Lemma~\ref{Lem:NecCond}}\label{App:NecCond}
This lemma can be proved by contradiction. First, assume that  $\{\l(\ell_n^{(s)}\r)^*, t_n^*, P_n^*\}$ is the optimal solution satisfying $\alpha_n \l(\ell_n^{(s)}\r)^*+\frac{t_n^*}{h_n} f\l(\frac{\l(\ell_n^{(s)}\r)^*}{R_n t_n}\r) < \eta P_n^* h_n T_0$. Then, there exists another feasible power allocation policy, denoted by $P_n^{(1)}$, satisfying  $P_n^{(1)}<P_n^{*}$ and $\alpha_n {(\ell_n^{(s)})}^*+\frac{t_n^*}{h_n} f\l(\frac{\l(\ell_n^{(s)}\r)^*}{R_n t_n}\r) = \eta P_n^{(1)} h_n T_0$. It leads to the following 
\begin{gather*}
\sum_{n=1}^N a_n \log(1+(\ell_n^{(s)})^* )-c\sum_{n=1}^N P_n^{(1)} T_0 >\\
 \sum_{n=1}^N a_n \log(1+(\ell_n^{(s)})^*)-c\sum_{n=1}^N P_n^* T_0,
\end{gather*}
which contradicts the optimality.  Similarly, if the optimal solution leads to $\beta_n \l(\ell_n^{(s)}\r)^*+t_n^* < T$, then there exist another feasible policy with $(\ell_n^{(s)})^{(1)}$ satisfying $(\ell_n^{(s)})^{(1)}=\frac{T-t_n^*}{\beta_n}>\l(\ell_n^{(s)}\r)^*$. This new policy yields larger operator's reward as
\begin{gather*}
\sum_{n=1}^N a_n \log(1+(\ell_n^{(s)})^{(1)})-c\sum_{n=1}^N P_n^* T_0 > \\
\sum_{n=1}^N a_n \log(1+\l(\ell_n^{(s)}\r)^*)-c\sum_{n=1}^N P_n^* T_0,
\end{gather*}
contradicting the assumption. Combing above together leads to the desired results.

\section{Proof of Lemma~\ref{Lem:conv}}\label{App:conv}
First, it is easy to prove that $\sum_{n=1}^N a_n \log\l(1+\frac{T-t_n}{\beta_n}\r)$ and $\frac{\alpha_n(T-t_n)}{\eta h_n \beta_n t_n}$ are concave functions for $t_n$. Next, the second derivative of function $E_n^t(t_n)$ is:
\begin{gather*}
\frac{\partial^2 E_n^t(t_n)}{\partial t_n^2}=\frac{N_0{(T\ln2)}^2}{\eta (h_n B R_n \beta_n)^2 t_n^3}2^\frac{T-t_n}{B R_n \beta_n t_n}>0
\end{gather*}
for $0<{t_n}<T$. Thus, $E_n^t(t_n)$ is a convex function and $-E_n^t(t_n)$ is a concave function. It follows that the objective function, being a summation of a set of convex functions, preserves the convexity. Combining it with the convex constraints yields the desired result.

\section{Proof of Corollary~\ref{Cor:StruTime}}\label{App:StruTime}
From \eqref{Eq:LosslessOpt}, the utility weight can be expressed as:
\begin{gather*}
a_n = \frac{(\lambda^*+c)N_0[\beta_n+(T-t_n^*)]}{\eta h_n^2} \times \\
\l[\l(\frac{T\ln2}{B R_n \beta_n t_n^*}-1\r)e^{\frac{T\ln2}{B R_n \beta_n t_n^*}-\frac{\ln2}{B R_n \beta_n}}+\frac{\alpha_n h_n}{N_0 \beta_n}+1\r].\\
\end{gather*}
Given $a_1<a_2$, it can be directly observed from the above equation that $t_1>t_2$. As for channel gain, it should satisfy the following equation:
\begin{gather*}
h_n \l[\frac{a_n \eta h_n}{(\lambda^*+c)N_0[\beta_n+(T-t_n^*)]}-\frac{\alpha_n}{N_0 \beta_n}\r] = \\
\l(\frac{T\ln2}{B R_n \beta_n t_n^*}-1\r)e^{\frac{T\ln2}{B R_n \beta_n t_n^*}-\frac{\ln2}{B R_n \beta_n}}+1.
\end{gather*}
Since the item $\l(\frac{T\ln2}{B R_n \beta_n t_n^*}-1\r)e^{\frac{T\ln2}{B R_n \beta_n t_n^*}-\frac{\ln2}{B R_n \beta_n}}+1$ and $N_0[\beta_n+(T-t_n^*)]$ are both decreasing with $t_n$, $t_1$ should be larger than $t_2$ given $h_1<h_2$, ending the proof.

\section{Proof of Proposition~\ref{Pro:LosslessPowerAllo}}\label{App:LosslessPowerAllo}
First, combing $\alpha_n \l(\ell_n^{(s)}\r)^*+\frac{t_n^*}{h_n} f\l(\frac{\l(\ell_n^{(s)}\r)^*}{R_n t_n}\r) = \eta P_n^* h_n T_0$ and $\beta_n \l(\ell_n^{(s)}\r)^*+t_n^*= T$ gives
\begin{equation*}
P_n^*=\frac{1}{h_n \eta T_0}\l[\frac{t_n^*}{h_n}f\l(\frac{T-t_n^*}{R_n \beta_n t_n^*}\r)+\frac{\alpha_n (T-t_n^*)}{\beta_n}\r].
\end{equation*}
Next, the first and second derivative of $P_n$ are given as follows.
\begin{gather*}
\frac{\partial P_n}{\partial t_n}\!=\!\frac{N_0}{\eta h_n^2 T_0}\l[\l(1-\frac{T\ln2}{B R_n \beta_n t_n}\r)2^\frac{T-t_n}{B R_n\beta_n t_n}\!-\!1\r]\!-\!\frac{\alpha_n}{\eta \beta_n h_n},\\
 \frac{\partial^2 P_n}{\partial t_n^2}\!=\!\frac{N_0{(T\ln2)}^2}{\eta T_0 (h_n B R_n \beta_n)^2 t_n^3}2^\frac{T-t_n}{B R_n \beta_n t_n}.
\end{gather*}
It can be observed that $\frac{\partial^2 P_n}{\partial t_n^2}>0$ for $0\le t_n\le T$. Thus $\frac{\partial P_n}{\partial t_n}$ is monotone increasing for $0\le t_n\le T$ and  $$\l(\frac{\partial P_n}{\partial t_n}\r)_{\rm{max}}=\frac{\partial P_n}{\partial t_n}\bigg|_{t_n=T}=-\frac{N_n \ln2}{\eta h_n^2 B R_n \beta_n}-\frac{\alpha_n}{\eta \beta_n h_n}<0.$$ It means $P_n$ is monotone decreasing with $t_n$ and  $({P_n})_{\rm{min}}={P_n}|_{t_n=T}=0$.
Combing  \eqref{Eq:KKT1} and $0\le t_n\le T$ gives 
\begin{align*}
\lambda^*
&>\frac{\frac{a_n}{\beta_n+(T-t_n^*)}}{\frac{\alpha_n}{\eta h_n \beta_n}-\frac{N_0}{\eta h_n^2}\l[\l(1-\frac{T\ln2}{B R_n \beta_n t_n^*}\r)2^\frac{T-t_n^*}{B R_n \beta_n t_n^*}-1\r]}-c\\
&=\frac{a_n \eta h_n}{q_n^{(r)}+q_n^{(s)}+q_n^{(c)} e^{\epsilon R_n}+\frac{N_0\ln2}{h_n B R_n}}-c = \phi_n.
\end{align*}
If $\lambda^*<\phi_n$, the optimal sensing time will exceed the time slot constraint, i.e., $t_n^* > T$, thus the sensor will not be selected and no power will be allocated to it, i.e., $P_n=0$, ending the proof.

\section{Proof of Proposition~\ref{Pro:LosslessDataComp}}\label{App:LosslessDataComp}
Based on \eqref{Eq:LosslessKKT3}, at least one of the two equations $\hat{\lambda}^*=0$ or $t_n^*+\frac{\ell_n^{(s)}\l(e^{\epsilon R_n^*}-e^{\epsilon}\r)}{f_n}+\frac{\ell_n^{(s)}}{s_n}-T=0$ should be satisfied. Assume that $t_n^*+\frac{\ell_n^{(s)}\l(e^{\epsilon R_n^*}-e^{\epsilon}\r)}{f_n}+\frac{\ell_n^{(s)}}{s_n}-T \neq 0$, then $\hat{\lambda}^*=0$ must hold, thus the optimal compression ratio $R_n^* \to \infty$ according to \eqref{Eq:LosslessKKT2}. However, the compression ratio cannot be infinite, which contradicts to the assumption, thus $t_n^*+\frac{\ell_n^{(s)}\l(e^{\epsilon R_n^*}-e^{\epsilon}\r)}{f_n}+\frac{\ell_n^{(s)}}{s_n}=T$ must be satisfied. The optimal solution should satisfy \eqref{Eq:LosslessKKT1}, i.e.,  
\begin{equation*}
\l(q_n^{(c)}+\frac{\hat{\lambda}^*}{f_n}\r)\epsilon e^{\epsilon R_n^*}-\frac{1}{h_n (R_n^{*})^2}f'\l(\frac{\ell_n^{(s)}}{t_n^* R_n^*}\r)=0.
\end{equation*}
Combining the above with the constraints  $\hat{\lambda}^*=-\dfrac{1}{h_n}g\l(\dfrac{\ell_n^{(s)}}{t_n^* R_n^*}\r)$ and $t_n^*=T-\dfrac{\ell_n^{(s)}\l(e^{\epsilon R_n^*}-e^{\epsilon}\r)}{f_n}-\dfrac{\ell_n^{(s)}}{s_n}$ according to \eqref{Eq:LosslessKKT2} and \eqref{Eq:LosslessKKT3} gives the desired result.

\section{Proof of Lemma~\ref{Lem:Losslessmonotone}}\label{App:Losslessmonotone}
The first derivative of function $z({R_n})$ is
\begin{gather*}
\frac{\partial z(R_n)}{\partial R_n}=\l[q_n^{(c)}-\frac{1}{h_n f_n}g\l(\frac{1}{d(R_n)R_n}\r)\r]\epsilon^2 e^{\epsilon R_n}+\\
\frac{N_0 \ln2}{B R_n^2 h_n}\l[\frac{2}{R_n}+\frac{\ln2[R_n d(R_n)]'}{B [R_n d(R_n)]^2}\l(1-\frac{R_n d'({R_n})}{d(R_n)}\r)\r]e^{\frac{\ln2}{B R_n d(R_n)}}.
\end{gather*}
First, it has $[R_n d({R_n})]^{'}=d({R_n})-R_n d'({R_n})$. Thus, $[R_n d({R_n})]'>0$ when $d({R_n})>R_n d'({R_n})$, and $[R_n d({R_n})]'<0$ when $d({R_n})<R_n d'({R_n})$. Combining them together leads to 
\begin{equation*}
\frac{\ln2[R_n d(R_n)]'}{B [R_n d(R_n)]^2}\l(1-\frac{R_n d'({R_n})}{d(R_n)}\r)>0, ~~\text{for}~ R_n\in[1, R_{\rm{max}}].
\end{equation*} Next, the first derivative of function $g(x)$ can be given as 
\begin{equation*}
g'(x)=-x f''(x)=-\frac{N_0\ln2^2}{B^2}2^{\frac{x}{B}}<0, ~~\text{for}~x>0.
\end{equation*}
It means that $g(x)$ is monotonically decreasing with $x$ and thus $g(x)\le g(x)|_{x=0}=0$ for  $x>0$. Since $\frac{1}{d(R_n)R_n}>0$, it has  $\l[q_n^{(c)}-\frac{1}{h_n f_n}g\l(\frac{1}{d(R_n)R_n}\r)\r]>0$ for $x>0$. Combining the above discussions leads to  $\frac{\partial z(R_n)}{\partial R_n}>0$ for $R_n\in[1, R_{\rm{max}}]$, ending the proof.

\section{Proof of Corollary~\ref{Cor:CompressionRatio}}\label{App:CompressionRatio}
\par Denote function $\phi(R_n)=q_n^{(c)} h_n$, according to equation~\eqref{Eq:LosslessR},
\begin{equation*}
\phi(R_n)\!=\!q_n^{(c)} h_n \!=\! \frac{1}{f_n}g\l(\frac{1}{d(R_n)R_n}\r)\!+\!\frac{1}{R_n^2 \epsilon e^{\epsilon R_n}}f'\l(\frac{1}{d(R_n)R_n}\r).
\end{equation*}
The derivation of $\phi(R_n)$ is
\begin{equation*}
-\frac{N_0 \ln2}{B R_n^2 \epsilon e^{\epsilon R_n}}\l[\frac{2\!+\!\epsilon}{R_n \epsilon e^{\epsilon R_n}}\!+\!\frac{\ln2[R_n d(R_n)]'}{B [R_n d(R_n)]^2}\l(1\!-\!\frac{R_n \epsilon e^{\epsilon R_n}}{f_n d(R_n)}\r)\r]e^{\frac{\ln2}{B R_n d(R_n)}}.
\end{equation*}
Since the term $\frac{\ln2[R_n d(R_n)]'}{B [R_n d(R_n)]^2}\l(1-\frac{R_n \epsilon e^{\epsilon R_n}}{f_n d(R_n)}\r)$ is always positive as proved in Appendix~\ref{App:Losslessmonotone}, the function $\phi(R_n)$ is monotonically decreasing for $R_n\in[1, R_{\rm{max}}]$. Given $q_1^c>q_2^c$, it can be observed that $R_1<R_2$, similarly $R_1<R_2$ when $h_1>h_2$, ending the proof.
\end{appendices}

\bibliographystyle{ieeetr}

\end{document}